\documentclass{article}

\usepackage[english]{babel}

\usepackage[letterpaper,top=2cm,bottom=2cm,left=3cm,right=3cm,marginparwidth=1.75cm]{geometry}

\usepackage{amsmath}
\usepackage{graphicx}
\usepackage[colorlinks=true, allcolors=blue]{hyperref}
\usepackage{natbib}

\title{The rotational kinetic Sunyaev-Zeldovich contribution to the temperature asymmetry toward the M31 halo}

\author{Noraiz Tahir$^{1,2,*}$; Francesco De Paolis$^{1,2,3,\dagger}$; Asghar Qadir$^{4,\ddagger}$; \\ Achille A. Nucita$^{1,2,3,\S}$\\
$^1$ Department of Mathematics and Physics ``E. De Giorgi'', \\ University of Salento, Via per Arnesano, I-73100  Lecce, Italy. \\
$^2$ INFN, Sezione di Lecce, Via per Arnesano, 73100 Lecce, Italy. \\
$^3$ INAF, Sezione di Lecce, Via per Arnesano, 73100 Lecce, Italy. \\
$^4$ Abdus Salam School of Mathematics, G.C. University, Lahore, Pakistan. \\
$^*$ noraiz.tahir@le.infn.it \\
$^\dagger$ francesco.depaolis@le.infn.it \\
$^\ddagger$ asgharqadir46@gmail.com \\
$^\S$ achille.nucita@le.infn.it \\
}

\begin{document}
\maketitle

\begin{abstract}
Temperature asymmetry in the cosmic microwave background (CMB) data by the {\it Planck} satellite has been discovered and analyzed toward several nearby edge-on spiral galaxies. It provides a way to probe galactic halo rotation, and to constrain the baryon fraction in the galactic halos. The frequency independence of the observed data provides a strong indication of the Doppler shift nature of the effect, due to the galactic halo rotation. It was proposed that this effect may arise from the emission of cold  gas clouds populating the galactic halos. However, in order to confirm this view, other effects that  might give rise to a temperature asymmetry in the CMB data, have to be considered and studied in detail. The main aim of the present paper is to estimate the contribution in the CMB temperature asymmetry data due to the free-free emission by hot gas (particularly electrons) through the rotational kinetic Sunyaev-Zeldovich (rkSZ) effect. We concentrate, in particular, on the M31 galactic halo and compare the estimated values of the rkSZ induced temperature asymmetry with those obtained by using the SMICA pipeline of the {\it Planck} data release, already employed to project out the SZ sources and for lensing studies. As an additional consistency check, we also verified that the hot gas diffuse emission in the X-ray band does not exceed that detected in the  soft X-ray band by {\it ROSAT} observations. We note that our results clearly show that the rkSZ effect gives only a minor  contribution to the observed M31 halo temperature asymmetry by {\it Planck} data.
\end{abstract}

\section{Introduction}
To describe the Universe we need to know what is in it and where all its components reside. This is a difficult task. For example, the ``missing baryon problem'' is one of the unsolved issues in modern cosmology. We know that $\sim 5\%$ of the Universe is composed of baryons \citep{ade2016baryons}, but not all these baryons have been detected as yet. What we can see is $\approx 60\%$ of the total baryonic budget constituting the luminous parts of the galaxies lying in what is called the  Layman $\alpha$ forest, in the warm-hot intergalactic gas (WHIG), in the the warm-hot intergalactic medium (WHIM) in clusters of galaxies, in the circumgalactic medium (CGM), and in the form of cold gas \citep{2003Ap&SS.284..697B,Cen_1999,li2018baryon, katherine1996baryon, gerhard1996baryonic}. The remaining part of the baryonic budget is still hidden. It has been suggested that galactic halos contain a non-negligible fraction of these baryons \citep{cen2006baryons,fraser2011estimate,nicastro2008missing,zhang2021measuring,2019IJMPD..2850088T,tahir2019seeing,2018MNRAS.476.5127R}, but in what form and how many baryons are present there are still open questions. 

The analysis of the cosmic microwave background (CMB) data of WMAP and {\it Planck} for various nearby edge-on spiral galaxies showed the existence of a large-scale (up to about 100 kpc in many cases)  temperature asymmetry (which means that one side turns out to be hotter than the other in the microwaves) with respect to the considered galaxy's rotational axis, with the important characteristic of being almost frequency independent. This gave a strong indication of a Doppler shift effect induced by the galactic halo rotation \citep{de2014planck, de2015planck, gurzadyan2015planck, de2016triangulum, gurzadyan2018messier, de2019rotating}. \citet{de1995scenario} proposed that galactic halos contain stable cold gas clouds in thermodynamic equilibrium with the CMB and that these clouds contribute a non-negligible fraction $f$ of the galactic halo dark matter. If these clouds are there, the only way to detect them is through the Doppler shift effect induced by the rotation of these clouds with the galactic halos \citep{de1995observing, de1998gamma}. The predicted Doppler shift was indeed detected  by the CMB data, and this data was then used to model the galactic halo rotation and estimate the gas cloud contribution \citep{2019IJMPD..2850088T, tahir2019seeing, qadir2019virial}. However, the main problem with the existence of these clouds is their stability \citep{padmanabhan1990statistical}, but the analysis by \citet{qadir2019virial} indicated that this equilibrium is possible and the clouds can be stable at the current CMB temperature.

It is clear that showing that these clouds may exist does not prove that they actually do exist and make up a substantial fraction of the missing baryons, nor that they are responsible for the temperature asymmetry in the CMB data observed toward several nearby galaxy halos. Indeed, before concluding that this in fact occurs it is necessary to analyze and exclude other effects that  may contribute, even partially, to the observed asymmetry toward nearby spirals. Many other effects, from a theoretical point of view, can contribute, such as the rotational kinetic Sunyaev-Zeldovich (rkSZ) effect by hot gas diffuse in the galactic halos \citep{PhysRevD.101.083016}, synchrotron emission by fast moving electrons \citep{dolag2000radio}, free-free (or bremsstrahlung) emission \citep{sun2010galactic}, and  anomalous microwave emission (AME) from dust grains in the halos \citep{leitch2013discovery}. The aim of the present paper is to make a first step forward in this direction and consider in detail the contribution of the rkSZ effect in the observed asymmetry in the {\it Planck} data toward M31. The contributions of the other effects will be considered elsewhere.

The manuscript is arranged as follows. In Section \ref{rkSZ} we model the density distribution of the hot gas in the M31 galaxy halo and its circular velocity. We therefore estimate the total mass of the hot gas component. In order to check the consistency of our adopted model we then estimate the diffuse X-ray flux toward the M31 halo and compare it with the diffuse emission observed in the soft X-ray band by the ROSAT satellite which is available \citep{1997MNRAS.287...10W} and in other works in the literature. We further proceed to estimate the expected temperature asymmetry values derived  from our modeling of the hot plasma in the M31 halo and compare them with the results obtained  using the SMICA pipeline of the  {\it Planck} data. Finally, our main conclusions are presented in Section \ref{results}.
\section{Modeling of the rkSZ effect \label{rkSZ}}
The Sunyaev-Zeldovich (SZ) effect has become a cornerstone of modern cosmology since it allows us to obtain a deeper understanding of several important  issues in astrophysics and cosmology. The SZ effect can be  divided into two types: the thermal SZ  (thSZ)  and the kinetic SZ  (kSZ) effects. The first  is due to the inverse Compton scattering of the CMB photons off the hot ICM \citep{zeldovich1969interaction}. The thSZ effect is redshift independent, so it is used to detect clusters of galaxies at high redshifts where other observational methods fail \citep{ade2015sz, bleem2020clustersurvey, hilton2021actszclusters}, and to determine the Hubble constant \citep{birkinshaw1999sunyaev, reese2002determining,mason2001measurement, kozmanyan2019hubble}. To be precise, the kSZ effect is due to the source's peculiar motion where the hot plasma  is responsible for the inverse Compton scattering on  CMB photons  \citep{sunyaev1980velocity}. This effect is often used to estimate the free electron distribution, and to infer the circumgalactic (CG) peculiar velocity once the free electron density distribution is given \citep{birkinshaw1999sunyaev}. The amplitude of the kSZ signal is sensitive to the product of the gas density and bulk velocity relative to the CMB frame, which is proportional only to the line-of-sight (los) component of the electron gas velocity in galaxies and galactic halos \citep{birkinshaw1999sunyaev}. However, the measurement of the kSZ effect, at the present level of accuracy, is extremely difficult. More precise experiments in the near future will enable us to study the kSZ effect in more detail.  The early detection of the kSZ effect was due to the proper motion of galaxy clusters. \citet{hand2012evidence} analyzed Atacama Cosmology Telescope (ACT) and   Baryon Oscillation Spectroscopic Survey (BOSS) data using mean pairwise statistics, which gave a new constraint to measure the kSZ effect. The same approach was then used by \citet{de2017detection} \& \citet{li2018measurement} for the case of galaxies. The kSZ signal due to the proper motion of clusters has   been detected in stacked data \citep{lavaux2013first,schaan2016evidence} and   through high-resolution imaging of various individual systems \citep{adam2017mapping}. It has also been detected in cross-correlation analyses of projected fields \citep{hill2016kinematic,2016PhRvD..94l3526F,dore2004beyond}. 

The kSZ effect induces two kinds of contributions: the first  derives from the motion of the source as a whole and appears as a monopole-like temperature shift centered on the source. For example, this contribution has been used frequently in the literature in order to trace the density profile of the diffuse gas in some galaxy clusters \citep{baxter2019constraining}. The  second contribution  derives from the internal motion of the hot gas within the source, and it may have a complicated morphology \citep{baldi2018kinetic}. In particular, the bulk rotational motion of the  gas is expected to give a dipole-like kSZ signal if one side of the source is moving toward the observer, while the other is moving away. This effect is referred to as the rotational kSZ, or rkSZ \citep{baxter2019constraining}. There is some agreement that the rkSZ effect may be particularly  active on  galaxy cluster scales \citep{cooray2002halo, chluba2002kinetic, baldi2018kinetic}, and in this case it may give rise to temperature asymmetry  on the order of $10~\mu$K. Recently, the rkSZ induced signal has been  detected at about $3\sigma$ confidence level toward about $10^4$ massive galaxies of low redshift with halo mass on the order of $10^{11}~M_{\odot}$, using the Planck data \citep{PhysRevD.101.083016}. 

\begin{figure}[ht]
   \centering
   \includegraphics[scale=.3]{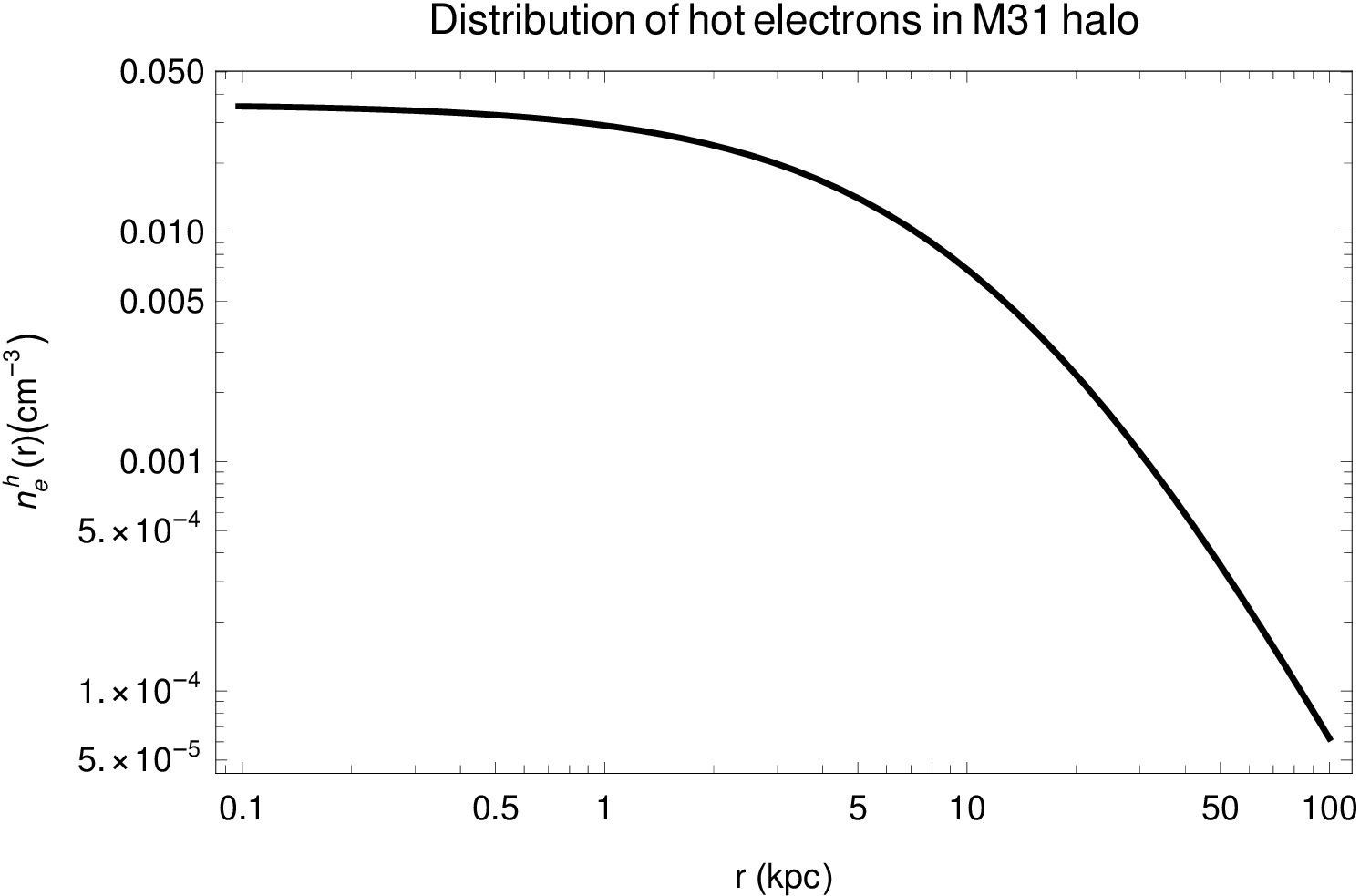}
   \caption{Hot electron gas number density profile in the M31 halo up to a galactocentric distance of about 100 kpc. The central value $n^h_{\circ}\simeq 3.4 \times 10^{-2}\,{\rm cm^{-3}}$  has been adopted.}
        \label{densityprofile}
    \end{figure}

As a first approximation and to get an idea of the induced effect in the CMB by the rkSZ effect, we consider the closest massive galaxy, M31. We assume the M31 halo to be filled by a hot plasma. Therefore,  the variation in the CMB temperature induced by the rkSZ effect depends on the  los integral of the hot gas number density and  on its circular velocity, and can be estimated through the equation
\begin{equation}
        \frac{\Delta T}{T}({\bf n})=\frac{\sigma_T}{c}\int_{los}n_e^h(r)~{\bf n}\cdot {\bf v}~dl,
    \label{eqnewdelta}
\end{equation}
where ${\bf n}$ is the unit vector that defines the point on the sky where the CMB temperature is measured, $\sigma_T$ is the Thomson cross section, $c$ is the speed of light, $n_e^h(r)$ is the hot electron number density distribution profile, $r$ is associated with the three-dimensional radial distance from the M31 center, and ${\bf v}$ is the hot gas velocity. Therefore,  eq. (\ref{eqnewdelta}) can be rewritten as
\begin{equation}
        \frac{\Delta T}{T} (R, \phi)=\frac{\sigma_T}{c}\int_{los} n_e^h(r)v_c(R) \cos\phi \sin i~dl,
        \label{deltat}
\end{equation}
where $v_c(R)$ is the circular velocity of the hot electron gas; $R$ is the projected radial distance, that is the distance with respect to the M31 rotation axis;  $\phi$ is the usual azimuthal angle; and $i=77.5^\circ$ is the inclination angle of M31  (i.e., the angle between the normal to the M31 disk and the line of sight). The hot gas is assumed to be at a temperature $\simeq 10^6$ K \citep{1997MNRAS.287...10W} and is taken to be responsible for the inverse Compton scattering of the CMB photons that   distort the background radiation from its blackbody spectral shape. Here we note that the step from eq. (\ref{eqnewdelta}) to eq. (\ref{deltat}) only requires a
spherically symmetric hot gas  distribution  moving along circular
orbits of radius $R$ with velocity $v(R)$. In our model, following  \citet{PhysRevD.101.083016}, the hot electron gas number density $n_e^h(r)$ is assumed to be given by
\begin{equation}
     n_e^h(r)= \frac{n_{\circ}^h}{\mu_e\left(\frac{r}{r_c}+\frac{3}{4}\right)\left(\frac{r}{r_c}+1\right)^2},
        \label{ne}
\end{equation}
where $n_{\circ}^h$ is the central hot electron number density, $\mu_e=1.18$ is the mean atomic weight per electron, and $r_c$ is the M31 halo core radius. In our  model we assume  $n_{\circ}^h \simeq 3.4 \times10^{-2}~{\rm cm^{-3}}$ and $r_c\simeq 15$ kpc \citep{1997MNRAS.287...10W}. The number density profile of the hot electron gas in the M31 halo, obtained   using eq. (\ref{ne}), is shown in Fig. \ref{densityprofile}.

\begin{figure}[ht]
        \centering
        \includegraphics[scale=.3]{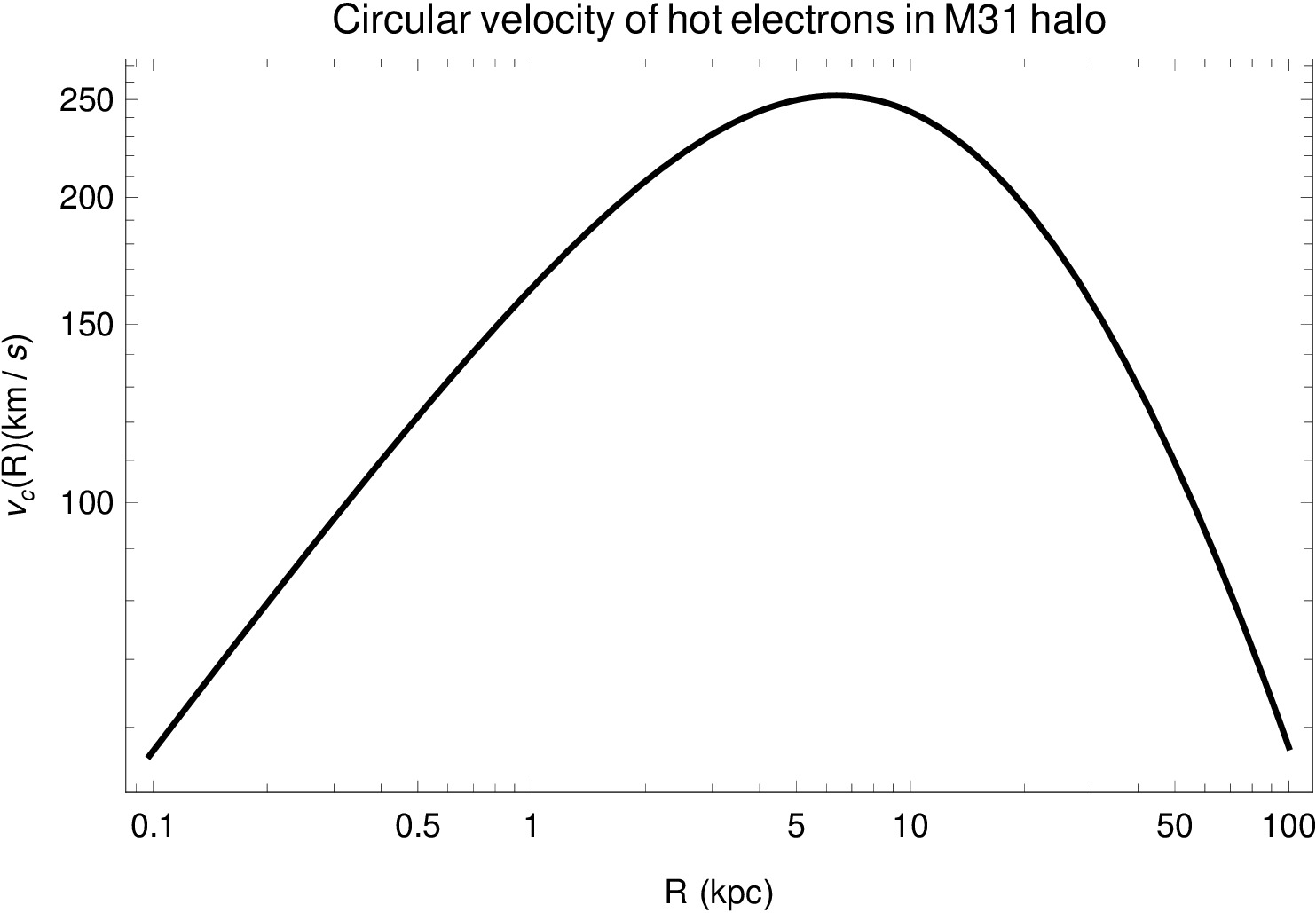}
        \caption{Hot gas circular velocity radial profile in the M31 halo up to a galactocentric distance of about 100 kpc.}
        \label{velocitydistribution}
\end{figure}

\citet{PhysRevD.101.083016} also give an explicit expression for the hot gas circular velocity profile $v_c(R)$ expressed as
\begin{equation}
        v_c(R)=\sqrt{\alpha\left[\ln \left(1+\frac{R}{R_{vir}}\right)-\left(\frac{R}{R_{vir}}\right)\left(1+\frac{R}{R_{vir}}\right)^{-1}\right]},
        \label{vc}
\end{equation}
where $\alpha=(4 \pi G r_c^2 m_p n_{\circ}^h R_{vir})/R$, $G$ is  Newton's gravitational constant, $m_p=1.67\times 10^{-24}$ g is the proton mass, and $R_{vir}$ is the M31 halo virial radius which is assumed to be $R_{vir}\simeq 200$ kpc \citep{tamm2012stellar}. The hot electron circular velocity radial profile  is shown in Fig. \ref{velocitydistribution}.

\begin{figure}[ht]
        \centering
        \includegraphics[scale=.3]{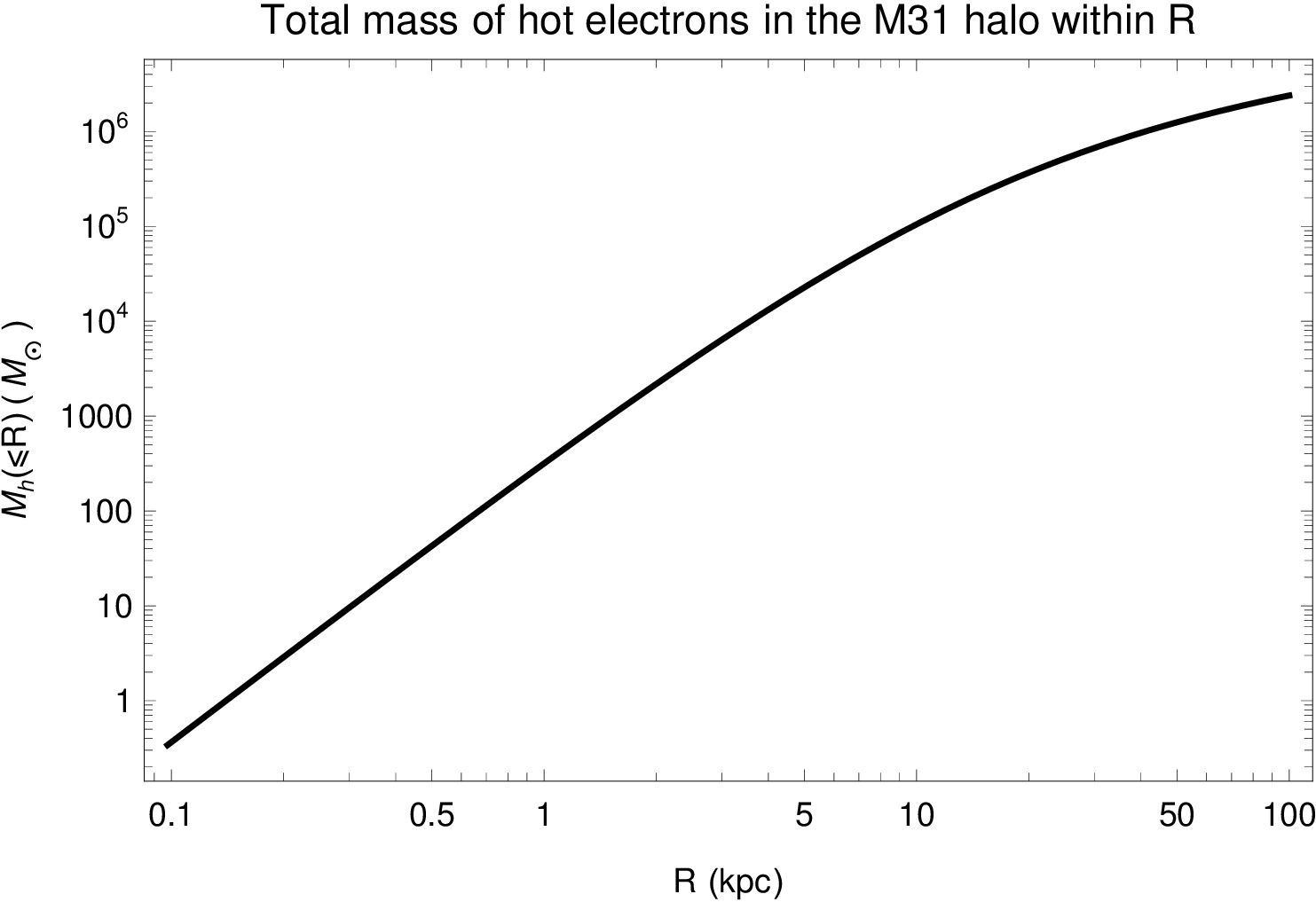}
        \caption{M31 halo hot gas mass within $R$ calculated through eq. (\ref{mass}).}
        \label{massdistribution}
\end{figure}

With the model above, the total hot gas mass in the M31 halo within $R$ can be easily calculated as
\begin{equation} 
        M_{h}(\leq R)=\int_{0}^{R}4\pi r^2 m_p n_e^h(r) dr.
        \label{mass}
\end{equation} 
In Fig. \ref{massdistribution} we give the estimated hot gas mass in the M31 halo up to about 100 kpc. It can be seen that our simple model returns a value for the hot gas mass in the M31 halo in agreement with that estimated in \citet{1997MNRAS.287...10W}. The estimated hot gas mass in \citet{1997MNRAS.287...10W} is $\simeq 7.0 \times 10^6~M_{\odot}$ up to 20 kpc, while from our model we get about $ 6.0 \times 10^6~M_{\odot}$ within 20 kpc and $\simeq 2.0 \times 10^7~M_{\odot}$ up to about 100 kpc.   
Our next aim is to  estimate, on the basis of the model above for the hot gas electrons radial profile, the X-ray flux emitted by the hot electron gas in the M31 halo in order to perform an additional consistency check of the model, by verifying that the diffuse flux does not exceed the observational data in the soft X-ray band  available in the literature.

As done before, we assume that the hot diffuse  gas in the halo of the M31 galaxy is at a temperature  $T\simeq 10^{6}$ K \citep{1997MNRAS.287...10W}. It should then give rise to a thermal emission through the bremsstrahlung emission, which should be responsible for a soft X-ray background. The emissivity of this plasma can be described, as a function of the galactocentric distance $R$ and the emission frequency $\nu$, by the relation  
\begin{align}
        \epsilon_{ff}(R,\nu)=&6.80 \times 10^{-32}Z (n_e^h(R))^2 T^{-0.5}\times \nonumber \\ &\exp\left(-\frac{h \nu}{k_B T}\right)G_{ff}(\nu)~{\rm erg~s^{-1} cm^{-3}~Hz^{-1}},
        \label{emissivity}
\end{align}
where $Z$ is the gas metallicity, assumed to be $\simeq 0.074$ \citep{2020AJ....160...41G}; $h$ is  Planck's constant; $k_B$ is the Boltzmann constant; and $G_{ff}$ is the Gaunt factor. In the relevant case of X-ray emission, $G_{ff}$ can be written as \citep{2011ihea.book.....R}
\begin{equation}
        G_{ff}(\nu)= \left(\frac{\sqrt{3}}{\pi}\right) \log\left[\frac{k_B T}{h \nu}\right].
        \label{gaunt}
\end{equation}
The total radiated power per unit volume by free-free emission from the hot electron gas in the M31 halo can be thus estimated by substituting eq. (\ref{gaunt}) in eq. (\ref{emissivity}) and integrating over the energy band which is assumed to be $0.5-2$ keV, which is the soft band of {\it ROSAT}. The total radiated power per unit volume can be written as
\begin{align}
        &P_{ff}(R)=\int_{\nu_i}^{\nu_f} \epsilon_{ff}(R,\nu)~d\nu =\nonumber \\
                 &\frac{4.57 \times 10^{-26}}{(0.75+2.65 \times 10^{-21}R)^2 (1+ 2.65 \times 10^{-21} R)^4}~{\rm erg~s^{-1}~cm^{-3}}.
                \label{power}
\end{align}
Thus, the total luminosity emitted within the radius $R$ can be estimated by
\begin{equation}
        \Gamma_{ff}(\leq R)=\int_{A}P_{ff}(R)~dA~{\rm erg~s^{-1}}, 
        \label{luminosity}
\end{equation}
where $A$ is the aperture toward M31 halo within R. Similarly, the flux density can be calculated by
\begin{equation}
        \Phi_{ff}(R)=\frac{\Gamma_{ff}(R)}{4\pi D^2}~{\rm erg~s^{-1}~cm^{-2}},
        \label{flux}
\end{equation}
where  $D \simeq 744$ kpc is the distance to the M31 center \citep{ribas2005first}. In Table 2 of \citet{1997MNRAS.287...10W}, the diffuse soft X-ray  flux for the energy band $0.5-2$ keV toward the M31 halo is estimated. Assuming a hot gas core radius $r_c= 15$ kpc, they derive the luminosity of the spherical halo $\approx 7.2 \times 10^{39}~{\rm erg~s^{-1}}$ up to about 20 kpc.
In order to  check the  consistency of our model, we compare the X-ray luminosity toward the M31 halo obtained by using our model results by using eq. (\ref{luminosity})
and then the flux density from eq. (\ref{flux}). Our estimated luminosity $\Gamma_{ff}$ comes out to be $\approx 2.80 \times 10^{40}~{\rm erg~s^{-1}}$ up to 20 kpc, in substantial agreement with that estimated from observations by \citet{1997MNRAS.287...10W} (see their Table 2). The corresponding X-ray flux $\Phi_{ff}$ turns out to be $\approx 4.24 \times 10^{-10}~{\rm erg~s^{-1}~ cm^{-2}}$. We can therefore conclude that our model results are in  good enough agreement with the observations.

\begin{figure}[ht]
        \centering
        \includegraphics[scale=.16]{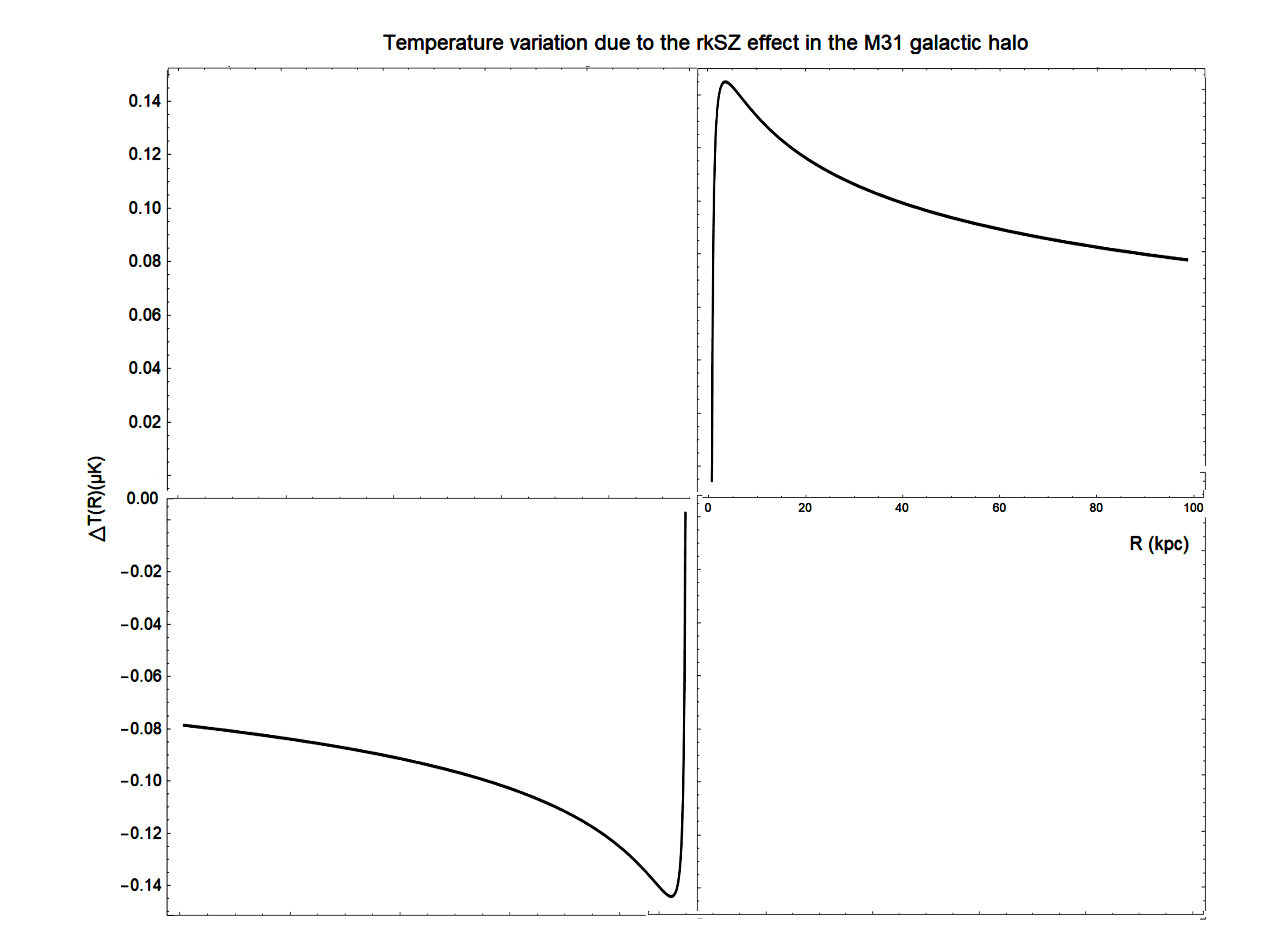}
        \caption{Maximum value of the rkSZ-induced temperature variation along the major axis of the M31 galactic halo as a function of the M31 halo projected radius $R$ is given.}
        \label{radialprofiledeltat}
\end{figure}

We are now in a position to estimate the temperature excess $\Delta T$ induced by the rkSZ effect, through eq. (\ref{deltat}). In Fig. \ref{radialprofiledeltat} we give the maximum value of the contribution from the rkSZ effect to the temperature variation of the CMB. This result is obtained by using eq. (\ref{deltat}) and scanning the major axis of the M31 halo from left to right. As  can be seen, the  rkSZ-induced temperature variation peaks at about 5 kpc from the M31 center and then slowly decreases toward the external M31 halo region. Similarly, in Table \ref{tab1} we give the maximum value of the estimated temperature variation induced by the rkSZ effect (central column) at various radial distances from the M31 center (specified in the first column) and compare them with the detected excess given in the third column.  The  values given in the third column correspond to the SMICA processed data. Here we note that the  SMICA data were chosen because they are less contaminated   than  the other available {\it Planck} bands. It is new {\it Planck} data product and has already been adopted, for example to search for SZ sources and in lensing studies \citep{Planck2018IV,Aghanim2020}. We also note  that the central  column of Table  \ref{tab1} refers to the maximum value of the rkSZ-induced temperature variation along the major axis of the M31 galactic halo edge-on map, without including the instrument beam smoothing. As   can be seen, the estimated values of $\Delta T$ are much lower with respect to the values observed by the {\it Planck} satellite. Hence, we can safely say  that the contribution induced by the rkSZ effect toward the M31 halo turns out to be negligible.

Until now we have assumed that the hot gas temperature in the M31 halo is $T\simeq 10^6$ K, and the hot electron central number density is $n_{\circ}^h \simeq 3.4 \times 10^{-2}~{\rm cm^{-3}}$. The question arises of what   changes in the values of $\Delta T$ induced by the rkSZ effect by varying these parameters, and requiring that the diffuse X-ray flux values estimated in Table 2 of \citet{1997MNRAS.287...10W} are not violated. We increased the value of the hot gas temperature  to a few times $\sim 10^7$ K, and decreased the value of the central number density   to $\sim 10^{-4}-10^{-5}~{\rm cm^{-3}}$. As a result we find that the estimated temperature variation induced by the rKSZ effect increased only negligibly for each considered M31 halo radius. This makes our results robust enough.
\begin{table}[ht]
\caption{Estimated maximum temperature variation (central column) induced by the rkSZ effect at various radial regions (indicated in the left column) along the major axis of the M31 halo.  Equation (\ref{deltat}) has been used and a hot gas temperature at $T\simeq 10^{6}$ K has been assumed. In the right  column the SMICA observed temperature variation toward each region is also given.}              
\centering                                      
\begin{tabular}{c c c c}          
\hline\hline                        
                Region & Estimated & Observed \\
                 $R$ &  $\Delta  T_{est}$ &  $\Delta  T_{obs}$ \\
                kpc  & $\mu$ K & $\mu$K \\
\hline
                 21.4  &$1.16 \times 10^{-1}$ & 2.66   \\
                31.1  & $1.06 \times 10^{-1}$ & 6.44 \\
                41.4  & $9.98 \times 10^{-2}$ & 17.6 \\
                51.8  & $9.54 \times 10^{-2}$ & 39.2 \\
                77.8  & $8.45 \times 10^{-2}$& 21.7 \\
                103.8  & $7.86 \times 10^{-2}$ & 21.4 \\                            \hline 
\end{tabular}
\label{tab1}
\end{table}

\section{Conclusions \label{results}}
Galactic halos are certainly much less studied with respect to galactic disks, and there are still many unanswered questions related to their dynamics. The {\it Planck} data of the CMB  opened up a new  window to probe the halo dynamics in detail, and the rotation of disks and halos of various nearby edge-on spiral galaxies has been investigated using these data. The independence of the CMB temperature asymmetry of the frequency toward the M31 halo constitutes a strong indication of its Doppler nature as a consequence of the M31 halo rotation. What is not well understood is whether the true cause of the Doppler shift is  fully due to virial clouds populating the M31 halo or if  there is a non-negligible contribution from other effects, such as the rkSZ effect, free-free emission by free electrons, synchrotron radiation by fast moving electrons, and/or anomalous microwave emission by dust grains.

In the present paper we estimated in particular the contribution to the observed temperature asymmetry toward the M31 halo by the rkSZ effect. We started our analysis by constraining the hot gas parameters in the M31 halo. Following \citet{1997MNRAS.287...10W} we assumed that the central number density of the hot electron gas is $\approx 3 \times 10^{-2}~{\rm cm^{-3}}$ and its temperature  is $\approx 10^6$ K.

We then used eq. (\ref{ne})  to estimate the density distribution of this hot gas in the M31 halo (see also Fig. \ref{densityprofile}). Similarly, the circular velocity $v_c(R)$ of this hot gas was estimated by using eq. (\ref{vc}) and the resulting radial profile was shown in Fig. \ref{velocitydistribution}. The obtained values for the hot gas circular velocity results were found to be in the range $50-250$ km/s. We then estimated the total mass distribution of this hot gas in the M31 halo, as shown in Fig. \ref{massdistribution}, and found a total hot gas mass $\approx 1.9 \times 10^{7} M_{\odot}$ up to about 100 kpc.

 The next step was  checking if the assumed parameters for the hot gas in the M31 halo give results that are not in contradiction with observations. In order to tackle this problem we considered the diffuse X-ray flux due to free-free emission by hot electrons in the M31 halo, and as a result we estimated theoretically the expected luminosity and flux in $0.5-2$ keV energy band. The estimated luminosity was $\approx 2.8 \times 10^{40}~{\rm erg~s^{-1}}$ up to about 20 kpc, which is in rather good agreement with that observed in the soft X-ray band by {\it ROSAT} (see Section \ref{rkSZ}).

The final step was  estimating the expected temperature asymmetry, $\Delta T$, induced by the rkSZ effect as a function of the projected radial distance $R$ in the M31 halo by using eq. \ref{deltat}. As is clear from Table \ref{tab1}, the contribution due to the rkSZ effect can be considered  negligible since it can produce an effect certainly less than about $1\%$ of the  temperature anisotropy detected in the SMICA maps of the Planck satellite toward the M31 halo. This result makes  the findings in De Paolis et al. (2011, 2014), and in particular the interpretation of the temperature anisotropy detected toward the M31 halo as being due to a population of cold clouds rotating about a rotation axis, more robust. It cannot be excluded, however, that future observations such as those expected in the 220 GHz channel (which is the cross-over frequency of the tSZ component where the kSZ signal is expected to be larger)  of the Atacama Cosmology Telescope (ACT) (see, e.g., \citealt{bleem2022}), of the  Simons observatory \citep{caimapo2022}, of the CMB-S4 experiment \citep{galli2022}, and  of CMB-HD \citep{sehgal2019cmb}, can bring more precise CMB data allowing a mapping of the rotation of galactic halos using the rkSZ effect.

\section*{Acknowledgements}
NT, FDP and AAN acknowledges the TAsP (Theoretical Astroparticle Physics) and Euclid INFN projects. The anonymous referee is also acknowledget for the useful and constructive report.


\begin{thebibliography}{}

    \bibitem[Adam et al. (2017)]{adam2017mapping} Adam, R., Bartalucci, I., Pratt, G.W. et al. 2017, A\& A, 598, A115
 
    \bibitem[Ade et al. (2015)]{ade2015sz} Ade, P.A.R., Aghanim, N., Arnaud, M. et al. 2015, A\& A, 594, A27
 
    \bibitem[Ade et al. (2016)]{ade2016baryons} Ade, P.A.R., Aghanim, N., Arnaud, M. et al. 2016, A\& A, 594, A13
  
    \bibitem[Aghanim et al. (2020)]{Aghanim2020} Aghanim, N., Akrami, Y., Ashdown, M. et al. 2020, A\&A, 641, A8
    
    \bibitem[Akrami et al. (2018)]{Planck2018IV} Akrami, Y., Ashdown, M., Aumont, J. et al. 2018, A\&A, 641, A4
 
    \bibitem[Baldi et al. (2018)]{baldi2018kinetic} Baldi, A.S., De Petris, M., Sembolini, F. et al. 2018, MNRAS, 479, 4028
  
    \bibitem[Baxter et al. (2019)]{baxter2019constraining} Baxter, E.J., Sherwin, B.D., \& Raghunathan, S. 2019, J. Cosmol. Astropart. Phys., 2019, 001
  
    \bibitem[Bleem et al. (2020)]{bleem2020clustersurvey} Bleem, L.E., Bocquet, S., Stalder, B. et al. 2020, ApJ, 247, 25 
  
    \bibitem[Bleem et al. (2022)]{bleem2022} Bleem, L.E., Crawford, T.M., Ansarinejad, B. et al. 2022, ApJS, 258, 36
  
    \bibitem[Birkinshaw (1999)]{birkinshaw1999sunyaev} Birkinshaw, M. 1999, PR, 310, 2
  
    \bibitem[Burkert (2003)]{2003Ap&SS.284..697B} Burkert, A. 2003, APSS, 284, 697
  
    \bibitem[Cen e al. (1999)]{Cen_1999} Cen, R., \& Ostriker, J. 1999, P., ApJ, 514, 1
  
    \bibitem[Cen et al. (2006)]{cen2006baryons} Cen, R., \& Ostriker, J.P. 2006, ApJ, 650, 560
  
    \bibitem[Chluba et al. (2002)]{chluba2002kinetic} Chluba, J., \& Mannheim, K. 2002, A\& A, 396, 419
  
    \bibitem[Cooray et al. (2002)]{cooray2002halo} Cooray, A., \& Sheth, R. 2002, Phys. Rep, 372, 1
  
    \bibitem[De Bernardis et al. (2017)]{de2017detection} De Bernardis, F., Aiola, S., Vavagiakis, E.M. et al. 2017, J. Cosmol. Astropart. Phys., 2017, 008 
  
    \bibitem[De Paolis et al. (1995a)]{de1995scenario} De Paolis, F., Ingrosso, G., Jetzer, Ph. et al. 1995, A\& A, 295, 567
    
    \bibitem[De Paolis et al. (1995b)]{de1995observing} De Paolis, F., Ingrosso, G., Jetzer, Ph. et al. 1995, A\& A, 299, 647
    
    \bibitem[De Paolis et al. (1998)]{de1998gamma} De Paolis, F., Ingrosso, G., Jetzer, Ph. et al. 1995, A\& A, 510, L103
  
    \bibitem[De Paolis et al. (2011)]{de2011possible} De Paolis, F., Gurzadyan, V.G., Ingrosso, G. et al. 2011, A\& A, 534, L8
  
    \bibitem[De Paolis et al. (2014)]{de2014planck} De Paolis, F., Gurzadyan, V.G., Nucita, A., A. et al. 2014, A\& A, 565, L3
  
    \bibitem[De Paolis et al. (2015)]{de2015planck} De Paolis, F., Gurzadyan, V.G., Nucita, A.A. et al. 2015, A\& A, 580, L8
  
    \bibitem[De Paolis et al. (2016)]{de2016triangulum} De Paolis, F., Gurzadyan, V.G., Nucita, A.A. et al. 2016, A\& A, 593, A57
  
    \bibitem[De Paolis et al. (2019)]{de2019rotating} De Paolis, F., Gurzadyan, V.G., Nucita, A.A. et al. 2019, A\& A, 629, A87
  
    \bibitem[Dor{\'e} et al. (2004)]{dore2004beyond} Dor{\'e}, O., Hennawi, J.F., \& Spergel, D.N. 2004, ApJ, 606, 46
  
    \bibitem[Dolag et al. (2000)]{dolag2000radio} Dolag, K., \& Enlin, T.A. 2000, A \& A, 362, 151
  
    \bibitem[Ferraro et al. (2016)]{2016PhRvD..94l3526F} Ferraro, S., Hill, J.C., Battaglia, N. et al. 2016, PRD, 94, 123526

    \bibitem[Fraser-McKelvie et al. (2011)]{fraser2011estimate} Fraser-McKelvie, A., Pimbblet, K.A., \& Lazendic, J.S. 2011, MNRAS, 415, 1961
  
    \bibitem[Galli et al. (2022)]{galli2022} Galli, S., Pogosian, L., Jedamzik, K. et al. 2022, PRD, 105, 023513
  
    \bibitem[Gerhard et al. (1996)]{gerhard1996baryonic} Gerhard, O., \& Silk, J. 1996, ApJ, 472, 34
  
    \bibitem[Gilbert et al. (2020)]{2020AJ....160...41G} Gilbert, K.M., Wojno, J., Kirby, E.N. et al. 2020, ApJ, 160, 41
  
    \bibitem[Gurzadyan et al. (2015)]{gurzadyan2015planck} Gurzadyan, V.G., De Paolis, F., Nucita, A.A. et al. 2015, A\& A, 582, A77
  
    \bibitem[Gurzadyan et al. (2018)]{gurzadyan2018messier} Gurzadyan, V.G., De Paolis, F., Nucita, A.A. et al. 2018, A\& A, 609, A131
  
    \bibitem[Hand et al. (2012)]{hand2012evidence} Hand, N., Addison, G.E., Aubourg, E. et al. 2012, PRL, 109, 041101
  
    \bibitem[Herv{\'i}as-Caimapo et al. (2022)]{caimapo2022} Herv{\'i}as-Caimapo, C., Bonaldi, A., Brown, M.L. et al. 2022, ApJ, 924, 11
  
    \bibitem[Hill et al. (2016)]{hill2016kinematic} Hill, J.C., Ferraro, S., Battaglia, N. et al. 2016, PRL, 117, 051301
  
    \bibitem[Hilton et al. (2021)]{hilton2021actszclusters} Hilton, M., Sif{\'o}n, C., Naess, S. et al. 2021, ApJ, 253, 3 
  
    \bibitem[Katherine et al. (1996)]{katherine1996baryon}  Katherine, F.G., \& Thomas, P.A. 1996, MNRAS, 281, 1133
  
    \bibitem[Kozmanyan et al. (2019)]{kozmanyan2019hubble} Kozmanyan, A., Bourdin, H., Mazzotta, P. et al. 2019, A\& A, 621, A34
  
    \bibitem[Lavaux et al. (2013)]{lavaux2013first} Lavaux, G., Afshordi, N., \& Hudson, M.J. 2013, MNRAS, 430, 1617
  
    \bibitem[Leitch et al. (2013)]{leitch2013discovery} Leitch, E.M., \& Readhead, A. 2013, Adv. Astron., 6, 352407
  
    \bibitem[Li et al. (2018a)]{li2018baryon} Li, J.T., Bregman, J.N., Wang, Q.D. et al. 2018, ApJ Letters, 855, L24
  
    \bibitem[Li et al. (2018b)]{li2018measurement} Li, Y.C., Ma, Y.Z., Remazeilles, M. et al. 2018, PRD, 97, 023514
  
    \bibitem[Manson et al. (2001)]{mason2001measurement} Mason, B.S., Myers, S.T., \& Readhead, A.C.S. 2001, ApJ Letters, 555, L11
  
    \bibitem[Matilla et al. (2020)]{PhysRevD.101.083016} Matilla, J.M.Z., \& Haiman, Z. 2020, PRD, 101, 083016
  
    \bibitem[Nicastro et al. (2008)]{nicastro2008missing} Nicastro, F., Mathur, S., \& Elvis, M. Science, 319, 55
  
    \bibitem[Nicastro et al. (2017)]{nicastro2017decade} Nicastro, F., Krongold, Y., Mathur, S. et al. 2017, Astron. Nachr., 238, 2
  
    \bibitem[Padmanabhan (1990)]{padmanabhan1990statistical} Padmanabhan, T. 1990, PR, 188, 285
  
    \bibitem[Qadir et al. (2019)]{qadir2019virial} Qadir, A., Tahir, N., \& Sakhi, M. 2019, PRD, 100, 043028
  
    \bibitem[Richards et al. (2018)]{2018MNRAS.476.5127R} Richards, E.E., Van Zee, L., Barnes, K.L. et al. 2018, MNRAS,  476, 5127
  
    \bibitem[Reese et al. (2002)]{reese2002determining} Reese, E.D., Carlstrom, J.E., Joy, M.  2002, ApJ, 581, 53
  
    \bibitem[Ribas et al. (2005)]{ribas2005first} Ribas, I., Jordi, C., Vilardell, F. et al. 2005, ApJ, 635, L37
  
    \bibitem[Rosswog et al. (2011)]{2011ihea.book.....R} Rosswog, S., \& Br{\"u}ggen, M. 2011, (Cambridge Univ. Press, Cambridge, UK)
  
    \bibitem[Sunyaev et al. (1980)]{sunyaev1980velocity} Sunyaev, R.A., \& Zeldovich, Ya.B. 1980, MNRAS, 190, 413
  
    \bibitem[Schaan et al. (2016)]{schaan2016evidence} Schaan, E., Ferraro, S., Vargas-Magana, M. et al. 2016, PRD, 93, 082002
  
    \bibitem[Sun et al. (2010)]{sun2010galactic} Sun, X.H., Reich, W. 2010, Res. Astron. Astrophys., 10, 1287
  
    \bibitem[Sehgal et al. (2019)]{sehgal2019cmb} Sehgal, N., Aiola, S., Akrami, Y. et al. 2019, BAAS, 51, 6
  
    \bibitem[Tahir et al. (2019a)]{2019IJMPD..2850088T} Tahir, N., De Paolis, F., Qadir, A. et al. 2019, IJMPD, 28, 1950088
  
    \bibitem[Tahir et al. (2019b)]{tahir2019seeing} Tahir, N., De Paolis, F., Qadir, A. et al. 2019, AJOM, 8, 193
  
    \bibitem[Tahir et al. (2021)]{tahi2021submittedepjc} Tahir, N., Qadir, A., Sakhi, M. et al. 2021, EPJC, 2021, 81, 1
  
    \bibitem[Tamm et al. (2012)]{tamm2012stellar} Tamm, A., Tempel, E., Tenjes, P. et al. 2012, A\& A., 546, 4
 
    \bibitem[West et al. (1997)]{1997MNRAS.287...10W} West, R.G., Barber, C.R., \& Folgheraiter, E.L. 1997, MNRAS, 287, 10
  
    \bibitem[Zeldovich et al. (1996)]{zeldovich1969interaction} Zeldovich, Ya.B., \&  Sunyaev, R.A. APSS, 4, 301
  
    \bibitem[Zhang et al. (2012)]{zhang2021measuring} Zhang, Y., Liu, R., Li, H. et al. 2021, ApJ, 911, 58
    

\end{thebibliography}
\end{document}